\documentclass[aps,prd,twocolumn,showpacs,superscriptaddress,floatfix]{revtex4}

\usepackage{graphicx}  % needed for figures
\usepackage{dcolumn}   % needed for some tables
\usepackage{bm}        % for math
\usepackage{amssymb}   % for math
\usepackage{epsfig}
\usepackage{subfigure}
\usepackage{psfrag}
\usepackage{xspace}
\usepackage{multirow}
\usepackage{lscape}
\usepackage{comment}% comment out areas
\usepackage{color}

\newcommand{\sigmattbar} {\mbox{\ensuremath{\sigma_{t\bar{t}}}}}
\newcommand{\ttbar}     {\mbox{\ensuremath{t\bar{t}}}}
\newcommand{\ppbar}     {\mbox{\ensuremath{p\bar{p}}}}
\newcommand{\dzero}     {D0}

\newcommand{\pt}        {\mbox{$p_T$\ }}
\newcommand{\met}       {\mbox{\ensuremath{\slash\kern-.55emp_{T}}}}
\newcommand{\metsig}    {\mbox{\ensuremath{\sigma_{\slash\kern-.4emp_{T}}}}}

\newcommand{\herwig}    {{\sc herwig}}
\newcommand{\pythia}    {{\sc pythia}}
\newcommand{\alpgen}    {{\sc alpgen}}
\newcommand{\geant}     {{\sc geant3}}
\newcommand{\mcatnlo}   {{\sc mc@nlo}}
\newcommand{\Z}         {\mbox{$Z/\gamma^\star$}}
\newcommand{\mt}	{\mbox{\ensuremath{m_t}}}

\def\Journal#1#2#3#4{{#1} {\bf #2}, #3 (#4)}

\def\NIM{Nucl.~Instrum.~Methods~Phys.~Res.}
\def\NIMA{\NIM~A}

\def\PLB{{Phys.~Lett.}~B}
\def\PRL{Phys.~Rev.~Lett.}

\begin{document}
%\modulolinenumbers[1]
%\linenumbers

%%%%%%%%%%%%%%%%%%%%%%%%%%%%%%%%%%%%%%%%%%%%%%%%%%%%%%%%%%

\newcommand{\lumi}   {5.4~$\rm fb^{-1}$}

% COMBINED RESULT

\newcommand{\result} {7.56}
\newcommand{\erpos}  {+0.63}
\newcommand{\erneg}  {-0.56}
\newcommand{\err}    {^{\erpos}_{\erneg}}
\newcommand{\fullresult} {\ensuremath{\result \err\,{\rm(stat+syst)}}\xspace}

% DILEPTON RESULT

\newcommand{\resultll} {7.36}
\newcommand{\erposll}  {+0.90}
\newcommand{\ernegll}  {-0.79}
\newcommand{\errll}    {^{\erposll}_{\ernegll}}
\newcommand{\fullresultll} {\ensuremath{\resultll \errll\,{\rm(stat+syst)}}\xspace}

% LJETS RESULT 

\newcommand{\resultlj} {7.90}
\newcommand{\erposlj}  {+0.78}
\newcommand{\erneglj}  {-0.69}
\newcommand{\errlj}    {^{\erposlj}_{\erneglj}}
\newcommand{\fullresultlj} {\ensuremath{\resultlj \errlj\,{\rm(stat+syst)}}\xspace}

%%%%%%%%%%%%%%%%%%%%%%%%%%%%%%%%%%%%%%%%%%%%%%%%%%%%%%%%%%
% the following line is for submission, including submission to the
% arXiv!!
\hspace{5.2in} \mbox{Fermilab-Pub-11-233-E}

\title{Measurement of the {\boldmath $t\bar{t}$} production cross section
  using dilepton events in {\boldmath $p\bar{p}$} collisions}
\vspace*{0.1cm}
%%%%%%%%%%%%%%%%%%%%%%%%%%%%%%%%%%%%%%%%%%%%%%%%%%%%%%%%%%

%
\affiliation{Universidad de Buenos Aires, Buenos Aires, Argentina}
\affiliation{LAFEX, Centro Brasileiro de Pesquisas F{\'\i}sicas, Rio de Janeiro, Brazil}
\affiliation{Universidade do Estado do Rio de Janeiro, Rio de Janeiro, Brazil}
\affiliation{Universidade Federal do ABC, Santo Andr\'e, Brazil}
\affiliation{Instituto de F\'{\i}sica Te\'orica, Universidade Estadual Paulista, S\~ao Paulo, Brazil}
\affiliation{Simon Fraser University, Vancouver, British Columbia, and York University, Toronto, Ontario, Canada}
\affiliation{University of Science and Technology of China, Hefei, People's Republic of China}
\affiliation{Universidad de los Andes, Bogot\'{a}, Colombia}
\affiliation{Charles University, Faculty of Mathematics and Physics, Center for Particle Physics, Prague, Czech Republic}
\affiliation{Czech Technical University in Prague, Prague, Czech Republic}
\affiliation{Center for Particle Physics, Institute of Physics, Academy of Sciences of the Czech Republic, Prague, Czech Republic}
\affiliation{Universidad San Francisco de Quito, Quito, Ecuador}
\affiliation{LPC, Universit\'e Blaise Pascal, CNRS/IN2P3, Clermont, France}
\affiliation{LPSC, Universit\'e Joseph Fourier Grenoble 1, CNRS/IN2P3, Institut National Polytechnique de Grenoble, Grenoble, France}
\affiliation{CPPM, Aix-Marseille Universit\'e, CNRS/IN2P3, Marseille, France}
\affiliation{LAL, Universit\'e Paris-Sud, CNRS/IN2P3, Orsay, France}
\affiliation{LPNHE, Universit\'es Paris VI and VII, CNRS/IN2P3, Paris, France}
\affiliation{CEA, Irfu, SPP, Saclay, France}
\affiliation{IPHC, Universit\'e de Strasbourg, CNRS/IN2P3, Strasbourg, France}
\affiliation{IPNL, Universit\'e Lyon 1, CNRS/IN2P3, Villeurbanne, France and Universit\'e de Lyon, Lyon, France}
\affiliation{III. Physikalisches Institut A, RWTH Aachen University, Aachen, Germany}
\affiliation{Physikalisches Institut, Universit{\"a}t Freiburg, Freiburg, Germany}
\affiliation{II. Physikalisches Institut, Georg-August-Universit{\"a}t G\"ottingen, G\"ottingen, Germany}
\affiliation{Institut f{\"u}r Physik, Universit{\"a}t Mainz, Mainz, Germany}
\affiliation{Ludwig-Maximilians-Universit{\"a}t M{\"u}nchen, M{\"u}nchen, Germany}
\affiliation{Fachbereich Physik, Bergische Universit{\"a}t Wuppertal, Wuppertal, Germany}
\affiliation{Panjab University, Chandigarh, India}
\affiliation{Delhi University, Delhi, India}
\affiliation{Tata Institute of Fundamental Research, Mumbai, India}
\affiliation{University College Dublin, Dublin, Ireland}
\affiliation{Korea Detector Laboratory, Korea University, Seoul, Korea}
\affiliation{CINVESTAV, Mexico City, Mexico}
\affiliation{FOM-Institute NIKHEF and University of Amsterdam/NIKHEF, Amsterdam, The Netherlands}
\affiliation{Radboud University Nijmegen/NIKHEF, Nijmegen, The Netherlands}
\affiliation{Joint Institute for Nuclear Research, Dubna, Russia}
\affiliation{Institute for Theoretical and Experimental Physics, Moscow, Russia}
\affiliation{Moscow State University, Moscow, Russia}
\affiliation{Institute for High Energy Physics, Protvino, Russia}
\affiliation{Petersburg Nuclear Physics Institute, St. Petersburg, Russia}
\affiliation{Instituci\'{o} Catalana de Recerca i Estudis Avan\c{c}ats (ICREA) and Institut de F\'{i}sica d'Altes Energies (IFAE), Barcelona, Spain}
\affiliation{Stockholm University, Stockholm and Uppsala University, Uppsala, Sweden}
\affiliation{Lancaster University, Lancaster LA1 4YB, United Kingdom}
\affiliation{Imperial College London, London SW7 2AZ, United Kingdom}
\affiliation{The University of Manchester, Manchester M13 9PL, United Kingdom}
\affiliation{University of Arizona, Tucson, Arizona 85721, USA}
\affiliation{University of California Riverside, Riverside, California 92521, USA}
\affiliation{Florida State University, Tallahassee, Florida 32306, USA}
\affiliation{Fermi National Accelerator Laboratory, Batavia, Illinois 60510, USA}
\affiliation{University of Illinois at Chicago, Chicago, Illinois 60607, USA}
\affiliation{Northern Illinois University, DeKalb, Illinois 60115, USA}
\affiliation{Northwestern University, Evanston, Illinois 60208, USA}
\affiliation{Indiana University, Bloomington, Indiana 47405, USA}
\affiliation{Purdue University Calumet, Hammond, Indiana 46323, USA}
\affiliation{University of Notre Dame, Notre Dame, Indiana 46556, USA}
\affiliation{Iowa State University, Ames, Iowa 50011, USA}
\affiliation{University of Kansas, Lawrence, Kansas 66045, USA}
\affiliation{Kansas State University, Manhattan, Kansas 66506, USA}
\affiliation{Louisiana Tech University, Ruston, Louisiana 71272, USA}
\affiliation{Boston University, Boston, Massachusetts 02215, USA}
\affiliation{Northeastern University, Boston, Massachusetts 02115, USA}
\affiliation{University of Michigan, Ann Arbor, Michigan 48109, USA}
\affiliation{Michigan State University, East Lansing, Michigan 48824, USA}
\affiliation{University of Mississippi, University, Mississippi 38677, USA}
\affiliation{University of Nebraska, Lincoln, Nebraska 68588, USA}
\affiliation{Rutgers University, Piscataway, New Jersey 08855, USA}
\affiliation{Princeton University, Princeton, New Jersey 08544, USA}
\affiliation{State University of New York, Buffalo, New York 14260, USA}
\affiliation{Columbia University, New York, New York 10027, USA}
\affiliation{University of Rochester, Rochester, New York 14627, USA}
\affiliation{State University of New York, Stony Brook, New York 11794, USA}
\affiliation{Brookhaven National Laboratory, Upton, New York 11973, USA}
\affiliation{Langston University, Langston, Oklahoma 73050, USA}
\affiliation{University of Oklahoma, Norman, Oklahoma 73019, USA}
\affiliation{Oklahoma State University, Stillwater, Oklahoma 74078, USA}
\affiliation{Brown University, Providence, Rhode Island 02912, USA}
\affiliation{University of Texas, Arlington, Texas 76019, USA}
\affiliation{Southern Methodist University, Dallas, Texas 75275, USA}
\affiliation{Rice University, Houston, Texas 77005, USA}
\affiliation{University of Virginia, Charlottesville, Virginia 22901, USA}
\affiliation{University of Washington, Seattle, Washington 98195, USA}
\author{V.M.~Abazov} \affiliation{Joint Institute for Nuclear Research, Dubna, Russia}
\author{B.~Abbott} \affiliation{University of Oklahoma, Norman, Oklahoma 73019, USA}
\author{B.S.~Acharya} \affiliation{Tata Institute of Fundamental Research, Mumbai, India}
\author{M.~Adams} \affiliation{University of Illinois at Chicago, Chicago, Illinois 60607, USA}
\author{T.~Adams} \affiliation{Florida State University, Tallahassee, Florida 32306, USA}
\author{G.D.~Alexeev} \affiliation{Joint Institute for Nuclear Research, Dubna, Russia}
\author{G.~Alkhazov} \affiliation{Petersburg Nuclear Physics Institute, St. Petersburg, Russia}
\author{A.~Alton$^{a}$} \affiliation{University of Michigan, Ann Arbor, Michigan 48109, USA}
\author{G.~Alverson} \affiliation{Northeastern University, Boston, Massachusetts 02115, USA}
\author{G.A.~Alves} \affiliation{LAFEX, Centro Brasileiro de Pesquisas F{\'\i}sicas, Rio de Janeiro, Brazil}
% ...
%
% visitor_addresses.tex                        6 April 2011
%  available symbols are:
%  $\ast, \dag, \ddag, \S, \P, $\|$, $\ast\ast$, \dag\dag, \ddag\ddag ,\#
%
\collaboration{The D0 Collaboration\footnote{with visitors from
%{alton}
$^{a}$Augustana College, Sioux Falls, SD, USA,
%{burdin}
$^{b}$The University of Liverpool, Liverpool, UK,
%{haas,partridge}
$^{c}$SLAC, Menlo Park, CA, USA,
%{hesketh}
$^{d}$University College London, London, UK,
%{luna-garcia}
$^{e}$Centro de Investigacion en Computacion - IPN, Mexico City, Mexico,
%{podesta-lerma}
$^{f}$ECFM, Universidad Autonoma de Sinaloa, Culiac\'an, Mexico,
and 
%{weber}
$^{g}$Universit{\"a}t Bern, Bern, Switzerland.
%{garcia-guerra}
%$^{?}$UPIITA-IPN, Mexico City, Mexico,
%{hooper}
%$^{?}$Visitor from Bradley University, Peoria, IL, USA.
%{kozminski}
%$^{?}$}Visitor from Lewis University, Romeoville, IL, USA.
%{deceased}
%$^{\ddag}$Deceased.
}} \noaffiliation
\vskip 0.25cm

\date{May 26th, 2011}

%%%%%%%%%%%%%%%% TEXT text %%%%%%%%%%%%%%%%%%%%%%%%%%%%%%%%%%%%%%%%%%
\begin{abstract}
We present a measurement of the \ttbar\ production cross section 
\sigmattbar\ in \ppbar\ collisions at $\sqrt{s} = 1.96$~TeV using \lumi\ 
of integrated luminosity collected with the \dzero\ detector.
We consider final states with at least two jets and two leptons ($ee$, $e\mu$, $\mu\mu$),
and events with one jet for the the $e\mu$ final state as well. 
The measured cross section  is
\sigmattbar=\fullresultll~pb.  
This result combined with the cross section measurement in 
the lepton + jets final state yields
\sigmattbar=\fullresult~pb, which agrees with the standard model expectation.
The relative precision of 8\% of this measurement 
is comparable to the precision of the latest theoretical calculations.

\end{abstract}


\maketitle 

%%%%%%%%%%%%%%%%%% INTRODUCTION %%%%%%%%%%%%%%%%%%
\section{Introduction}
The precise measurement of the top quark pair (\ttbar)
production cross section (\sigmattbar) and its comparison
with the theoretical predictions provide important tests of perturbative
quantum choromodynamics (QCD).
At present, the most precise predictions of \sigmattbar\ are 
given by approximate next to next-to-leading order (NNLO) 
calculations~\cite{SMtheory_A,SMtheory_M,SMtheory_K}, 
that set a goal for the experimental
precision of the \sigmattbar\ measurement of (6 -- 9)\%.
Furthermore, because  \sigmattbar\ depends on the
top quark mass ($m_t$), it can be used to measure 
$m_t$~\cite{d0-topmass-fromxsection,d0-top-charged-higgs}.
Comparing the standard model (SM) prediction with the measured \sigmattbar\
value allows testing for the presence of physics beyond the SM, 
for instance, scenarios
in which the top quark would decay into a charged Higgs 
boson and a $b$~quark~\cite{d0-top-charged-higgs}.

In this Letter we present an updated measurement of \sigmattbar\ in
\ppbar\ collisions at $\sqrt s = 1.96$~TeV in the
dilepton ($\ell\ell^\prime$,  $\ell=e,\mu$) channel. 
Within the SM, top quarks decay almost 100\% of the time into a $W$ boson and a 
$b$ quark. 
We consider the leptonic decays of both $W$ bosons from top quark decay into 
$e\nu_e$, $\mu\nu_{\mu}$, or $\tau\nu_{\tau}$ 
(throughout this letter, $e$,  $\mu$, $\tau$
refer to both charge conjugate states: $e^\pm$, $\mu^\pm$ or $\tau^\pm$),  
where only leptonic decays of the
$\tau$ are considered.
%For $W\to\tau\nu_\tau$  decays, we consider only leptonic $\tau$ decays, 
%$\tau\to e \nu_e \nu_{\tau}$, $\tau\to \mu \nu_{\mu} \nu_{\tau}$.

This measurement complements the \sigmattbar\ measurements
in the lepton+jets ($\ell j$)  channel, in which  one of the $W$~bosons
 decays hadronically into a $q \bar{q}^{\prime}$ pair and the other
$W$ boson decays leptonically~\cite{d0-top-ljets-5.3fb-1,d0-top-ljets},
%performed with datasets corresponding to
%5.3~fb$^{-1}$ or lower of integrated luminosities 
as well as measurements
in the all-hadronic channel, in which both $W$~bosons decay
hadronically~\cite{d0-top_alljets-1fb-1}.
%performed with datasets corresponding to
%1.0~fb$^{-1}$ or lower of integrated luminosities

The measurement is based on data collected with the \dzero\ detector during Run~II
of the Fermilab Tevatron Collider that correspond to an integrated luminosity of 
$5.4 \pm0.3$~fb$^{-1}$. 
This result supersedes our previous measurement~\cite{d0-top-dilepton-1fb},
which used a dataset five times smaller than the one considered here.
The CDF Collaboration has measured \sigmattbar\ in the $\ell\ell^\prime$
channel using 2.8~fb$^{-1}$ of integrated luminosity~\cite{CDF-dilepton}.
The ATLAS and CMS Collaborations recently published their first  \sigmattbar\
measurements in $pp$ collisions at ${\sqrt s} = 7$~TeV~\cite{atlas-top,cms-top}.

%%%%%%%%%%%%%%%%%% DETECTOR %%%%%%%%%%%%%%%%%%
The \dzero\ detector is described in detail in~\cite{d0-detector}. The region of
the \dzero\ detector closest to the interaction region contains a tracking system
consisting of a silicon microstrip tracker and a central fiber tracker,
both located inside a superconducting solenoid magnet which generates a magnetic 
field of 2~T. Hits in 
these two detectors are used to reconstruct tracks from charged particles
in the pseudorapidity region $| \eta | <3$~\footnote{The pseudorapidity is 
defined as $\eta = - \ln [\tan(\theta/2)]$, where $\theta$ is the polar 
angle with respect to the proton beam.}.
%The SMT provides
%the capability to reconstruct the $p \bar p$ interaction vertex (PV) with a
%precision of about 40~$\mu$m in the place transverse to the beam axis and to
%determine the impact parameter of any track with respect to the PV with a precision
%between 20 and 50~$\mu$m, depending of the number of Hits in the SMT. 
Surrounding
the two tracking subdetectors are liquid-argon uranium calorimeters, segmented into
electromagnetic and hadronic sections. The central section of the calorimeter (CC) covers
pseudorapidities $| \eta | < 1.1$, and the  two end calorimeters (EC)  extend coverage
to  $| \eta | \approx 4.2$ with all three housed in separate cryostats. The muon system
surrounds the calorimeter and consists of three layers of tracking detectors and
scintillator trigger counters covering  $| \eta | < 2$. A toroidal iron magnet with a field of 1.8~T 
is located
outside the innermost layer of the muon detector. The luminosity is calculated from the rate
of  inelastic $p \bar p$ collisions measured with plastic scintillator arrays located
in front of the EC cryostats~\cite{lumi}.

The \dzero\ trigger is based on a three-level pipeline system. The first level 
is implemented in custom-designed hardware. 
The second level uses high-level processors to combine information from the different sub-detectors
to construct simple physics objects. The software-based third level uses full event information
obtained with a simplified reconstruction algorithm.

%In 2006, the \dzero detector was upgraded with a new calorimeter trigger and a new
%inner layer added to the silicon microstrip tracker~\cite{d0-l0}. We split the 
%data into two samples: Run IIa before this upgrade and Run IIb after it. 
%The corresponding integrated luminosities are 1.08~fb$^{-1}$ and  4.28~fb$^{-1}$
%respectively.

%%%%%%%%%%%%%%%%%% OBJECT ID %%%%%%%%%%%%%%%%%%
\section{Object Identification}
\label{sec:objectid}
The \ttbar\ dilepton final state contains two leptons (electrons, muons or an electron and a muon), 
at least two jets, and significant missing transverse momentum (\met) from escaping neutrinos. 

%These object reconstructed and identify at \dzero\
%as the following.

Electrons are identified as energy clusters with radius ${\cal R}=\sqrt{(\Delta\eta)^2+(\Delta\phi)^2}<0.2$ 
in the calorimeter ($\phi$ is the azimuthal angle)   which are
consistent in their longitudinal and transverse profiles
with those of an electromagnetic shower.
More than 90\% of the energy of the electron candidate must be deposited  in the 
electromagnetic part of the calorimeter, and less than 20\% of its energy may be deposited in an
annulus of $0.2 < {\cal R} < 0.4$ around its direction.
This cluster has to be matched to an inner detector track.
We consider electrons in the CC with  $|\eta| <1.1$ and in the EC with $1.5 < |\eta| < 2.5$.
In addition, we make a requirement on the electron likelihood discriminant, based on tracking 
and calorimeter information, at a cut value chosen to have
a high selection efficiency (near  85\%) for electrons, 
and good rejection (near 90\%) for jets misidentified as electrons.
Electrons fulfilling all these criteria are called "tight electrons". 

A muon is identified as a segment in at least one layer of the muon system
in the full acceptance of the  muon system that is
matched to a track in the central tracking system. 
Reconstructed muons must satisfy two isolation criteria.
First, the transverse energy deposited in an annulus
around the muon $0.1 < {\cal R} < 0.4$ ($E_T^{\mu,\text{iso}}$)
has to be less than 15\% of the transverse momentum of the muon ($p^{\mu}_T$).
Second, the sum of the transverse momenta of the tracks in a cone of radius ${\cal R}<0.5$ around the muon track 
in the central tracking system ($p_T^{\mu,\text{iso}}$) has to be less than 15\% of $p^{\mu}_T$.
Muons that fulfill these isolation criteria are referred to as "tight isolated muons".

Monte Carlo (MC) generated events are processed through a \geant~\cite{geant} based simulation of the D0 detector 
and the same reconstruction programs used for the data.
To simulate the effects from additional \ppbar\ interactions, zero bias events with no
trigger requirements selected randomly in collider data are overlayed on the fully simulated MC events.
Residual differences between data and MC simulation in the electron and muon 
\pt resolutions and identification efficiencies are corrected.
These corrections are derived from a sample of
\Z$\to\ell\ell$ events in data and MC, applying tight requirements
on one of the two leptons for selecting the events and using
the other one to measure the efficiencies and resolutions.

Jets are identified with a cone algorithm with radius ${\cal R}<0.5$~\cite{RunIIcone} in the range $|\eta|< 2.5$.
A jet energy scale correction (JES) is determined by calibrating
the energy deposited in the jet cone  
using transverse momentum balance in $\gamma$+jet and dijet events.
If a muon overlaps with the jet cone, 
the momentum of that muon is added to the jet \pt, assuming
that the muon originates from a semileptonic decay of a hadron belonging to the jet.

We require that the jets be matched to
at least two tracks originating from the vertex of the primary $p\bar{p}$ interaction (PV).
Jets in MC are corrected 
for the residual differences between data and MC in the energy resolution and JES.
These correction factors are measured by comparing data and MC 
in (\Z$\to ee$)+jets events.

We use a neural-network  (NN) tagging algorithm~\cite{btagging} to identify jets from $b$ quarks. The algorithm 
combines information from the  impact parameters of the tracks and variables 
that characterize the presence and properties of
secondary vertices within the jet in a single discriminant.
In order to use this information for $b$~tagging,
the jet is required to be matched to a jet built from tracks. 
Jets fulfilling this requirement are called taggable jets.
The NN discriminant has a value close to one for the b quark jets and close to zero for the 
light quark and gluon jets.

The \met\ is reconstructed from the energy deposited in the calorimeter 
cells. Corrections for lepton and jet $p_T$ are propagated into the \met.
The missing transverse momentum significance (\metsig) 
is defined in each event as a likelihood discriminant constructed using the ratio
of \met\ to its uncertainty.
%\metsig $=$ \met$/\sigma$,
%where $\sigma$ is the quadratic sum of the \met\ resolutions for jets, muons and electrons
%detected in the event.

More details about object identification can be found in~\cite{d0-top-dilepton-400pb}.

%%%%%%%%%%%%%%%% SELECTION, BACKGROUND %%%%%%%%%%%%%%%%%%

\section {Event Selection and Background Estimation}
\label{sec:selection}
The main sources of background in the $\ell\ell^\prime$ channel come from Drell-Yan and $Z$ boson production
(\Z$\to\ell\ell$), diboson production ({\sl WW, WZ, ZZ }), and instrumental background.
The instrumental background mainly arises from multijet and $(W \to \ell \nu)$+jets events in which one or two 
jets are misidentified as electrons and/or 
muons originating from the semileptonic decay of a heavy flavor hadron.
%appear isolated.

For this analysis we consider events that passed at least one of a set of single lepton triggers for the
$ee$ and $\mu\mu$ channels. For the $e\mu$ channel, we 
consider events selected by 
a mixture of single and multilepton triggers and lepton+jet
triggers.
%(mainly by the single leptons and electron-muon set of triggers). 
Efficiencies for single lepton triggers have been measured with
\Z$\to\ell\ell$ data. These efficiencies are found to be around 99\% for the $ee$ channel and 80\%
for $\mu\mu$. For the $e\mu$ channel the trigger efficiency is close to 100\%.

In order to separate  \ttbar\ signal events from background, the following selection
is applied:

\begin{itemize}

\item We require at least one PV in the beam interaction region with $ |z| < 60$ cm, where
$z$ is the coordinate along the beam axis,  and $z=0$ in the center of the detector.
At least three tracks must be associated with this PV.

\item We require at least two isolated leptons with $\pt>15$~GeV,
originating from the same PV, i.e., the difference between the $z$ coordinates 
of the two lepton tracks should be less than 2~cm, where the $z$ coordinate is calculated at the 
point of the track's closest approach to the beam. 

\item We select the two highest \pt\ leptons with opposite charges. 
%For the instrumental background determination we also use 
%events where both leptons have the same charge. We will refer to these
%events as the same sign sample.

\item In the $e\mu$ final state, we require the distance between the electron and the muon directions 
to be $R(e,\mu)>0.3$ to reduce the background from bremsstrahlung.
%\Z$\to\mu\mu\gamma$ 

\item In the $e\mu$ channel, we consider events with at least one jet with $\pt>20$~GeV.
In the $ee$ and $\mu\mu$ channels, we require at least two jets with $\pt>20$~GeV.

\item 
To further improve the signal purity of the selected sample, we apply additional 
selection criteria based on global event properties.
In the $e\mu$ channel with exactly one jet, we require $H_T>105$~GeV, 
where $H_T$ is the scalar sum of the transverse momenta of the leading lepton and the two leading jets.
In the $e\mu$ final state with two jets, we require $H_T>110$~GeV.
In the $ee$ final state, we require \metsig$>5$,
while in the $\mu\mu$ channel we require  \met$>40$~GeV and \metsig$>5$.
\end{itemize}

In order to estimate the signal efficiency and the background contamination,
we use the MC simulation for all contributions but for the instrumental background, 
the latter being derived from data.
The \ttbar\ and \Z\ events are generated with the tree level 
matrix element generator \alpgen ~\cite{Alpgen}
interfaced with the \pythia~\cite{Pythia} generator for parton 
showering and hadronization. 
Diboson events are generated with \pythia.
All simulated samples are generated using the CTEQ6L1 parton distribution functions~(PDFs)~\cite{Nadolsky:2008zw}.
The \Z\ samples are normalized to the NNLO cross section 
computed with the {\sc fewz} program~\cite{fewz}.
We separately  simulate \Z\ with heavy flavor (HF) quarks,  \Z$ + b\bar{b}$ (or \Z$+c\bar{c}$), using \alpgen\ 
and enhance the corresponding leading order cross sections by a factor 
%of~1.5 (1.7) 
estimated with the {\sc mcfm} program~\cite{Ellis:2006ar}.
The diboson samples are normalized to the next-to-leading order cross section calculated with {\sc mcfm}.
Uncertainties in these normalization factors are taken into account as systematic uncertainties.
In addition, we apply a correction to the $\Z$+jets simulation based on data to address
the imperfect modeling of the $Z$ boson \pt in the MC~\cite{zptrw}.
%These corrections are determined by comparing data and MC 
%distributions for the \Z$+jets\to\ell\ell$ dominated selections.

The instrumental background is estimated directly from data.
In the $ee$ and $e\mu$ channels we determine the contribution of events
with jets misidentified as electrons using the signal data sample but without
the electron likelihood discriminant requirement.
We extract the number of events with jets misidentified as electrons, $n_{f}$, and 
the number of events with real electrons, $n_e$,
by maximizing the function of the electron likelihood
distribution
\begin{equation}
\mathcal{L} = \prod_{i=1}^N [n_e S(x_i) + n_{f} B(x_i)] \frac{e^{- (n_e+n_f)}}{N!}\ ,
\end{equation}
where $N$ is the number of selected events, 
$x_i$ is the electron likelihood discriminant value in the event $i$,
and $ S(x_i)$ and $B(x_i)$ are the signal and background 
probability density functions (pdfs).
The signal pdf  is measured in \Z$\to ee$ data events.
The background pdf is measured in $e\mu$ events with the same selection as the analysis sample
but inverting the opposite sign lepton requirement 
(i.e., requiring leptons of the same sign)  without any topological requirement
but using muon with reversed isolation requirements: 
$E^{\mu,\text{iso}}_T/p^{\mu}_T > 0.2$ and $p^{\mu,\text{iso}}_T/p^{\mu}_T > 0.2$.
The total number of events with a jet misidentified as an electron 
is given by $n_f$ scaled for the integral of $B(x)$  over the region 
satisfying the likelihood requirement.
The estimation is performed separately in the CC and EC.
We find that the contribution of instrumental background to 
the $ee$ channel is negligible.
The limited size of the sample used to measure the background pdf is the main contribution to the systematic uncertainty on 
the instrumental background.
Together with a possible $p_T$ dependence of
the signal and background pdfs, these sources  typically  
lead to systematic uncertainties of about 50\%.

We also determine the number of events 
with an isolated muon arising from jets 
in the $e\mu$ and $\mu\mu$ channels.
This number is estimated as $n^{\mu}_{f} = N_{\text{loose}} f_{\mu}$, 
where $N_{\text{loose}}$ is the number of events 
in the same sign sample with loose isolation criteria on the muon:
$E^{\mu,\text{iso}}_T/p^{\mu}_T < 0.5$ and $p^{\mu,\text{iso}}_T/p^{\mu}_T < 0.5$,
and $f_{\mu}$ is the misidentification rate for isolated muons.
In the $\mu\mu$ final state, we apply these loose isolation criteria only
to one randomly chosen muon.
In the $e\mu$ channel, the number of events with jets misidentified as electrons
in the same sign sample is subtracted from $N_{\text{loose}}$.
The misidentification rate,
$f_{\mu}$, is determined in a dimuon sample with at least one jet.
In this sample we require one muon to be close to the jet ($R(\mu, \text{jet}) < 0.5)$
with reversed isolation criteria
$E^{\mu,\text{iso}}_T/p^{\mu}_T > 0.15$ and $p^{\mu,\text{iso}}_T/p^{\mu}_T > 0.15$.
The other muon defined as the probe, should pass the loose isolation criteria
$E^{\mu,\text{iso}}_T/p^{\mu}_T < 0.5$ and $p^{\mu,\text{iso}}_T/p^{\mu}_T < 0.5$.
We compute $f_{\mu}$ as the ratio of the number of events 
in which the probe muon passes
the tight isolation criteria to the total number of events in this same sign sample. 
The systematic uncertainty on the $f_{\mu}$ determination is about 10\% and results mainly
from the statistical uncertainty due to the limited size of the sample used for the muon misidentification
rate calculation and the potential dependence of the misidentification rate on $p_T$ and \met .

\begin{figure*}[!htb]
\includegraphics[width=.4\textwidth]{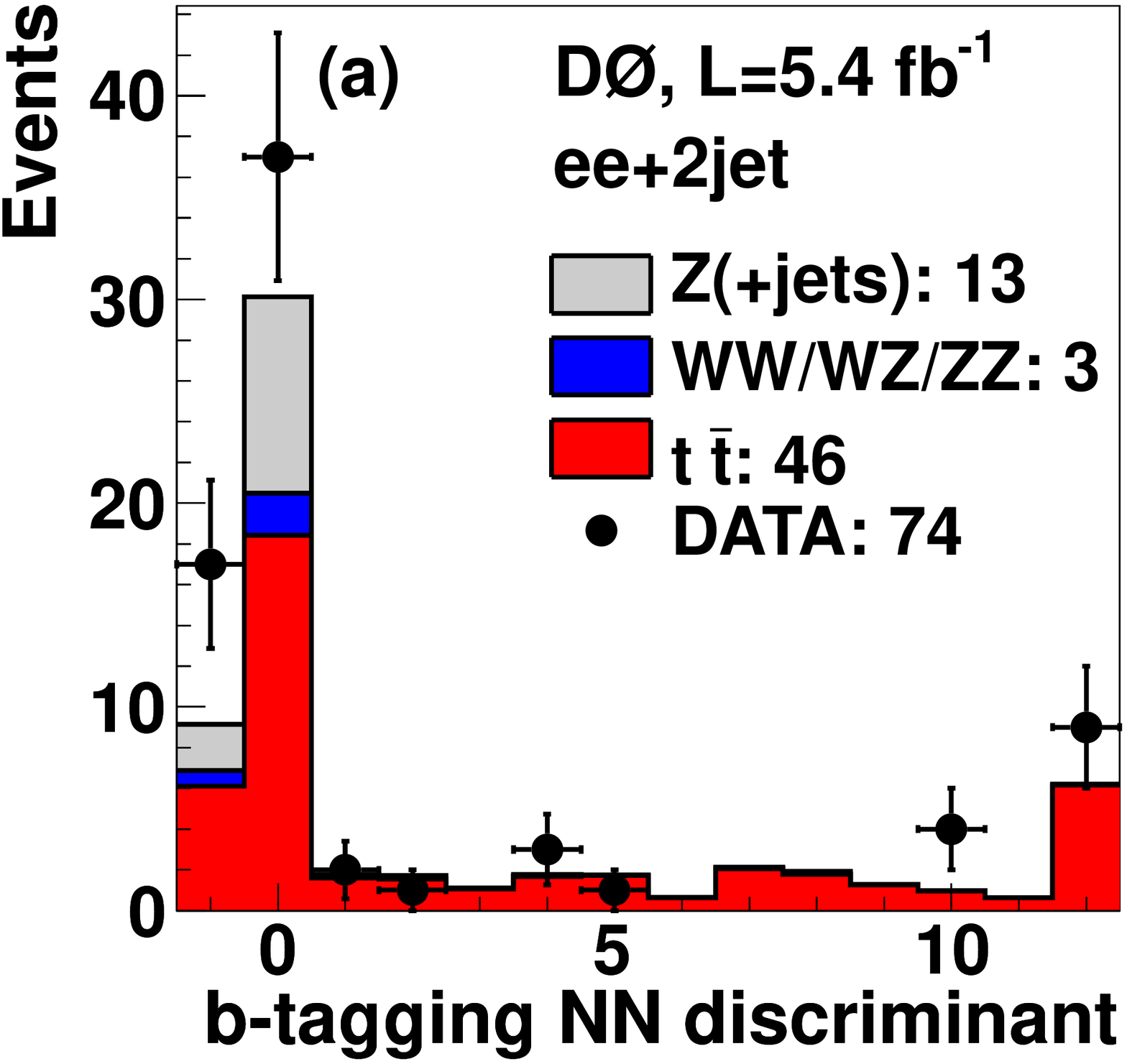}
\hfill
\includegraphics[width=.4\textwidth]{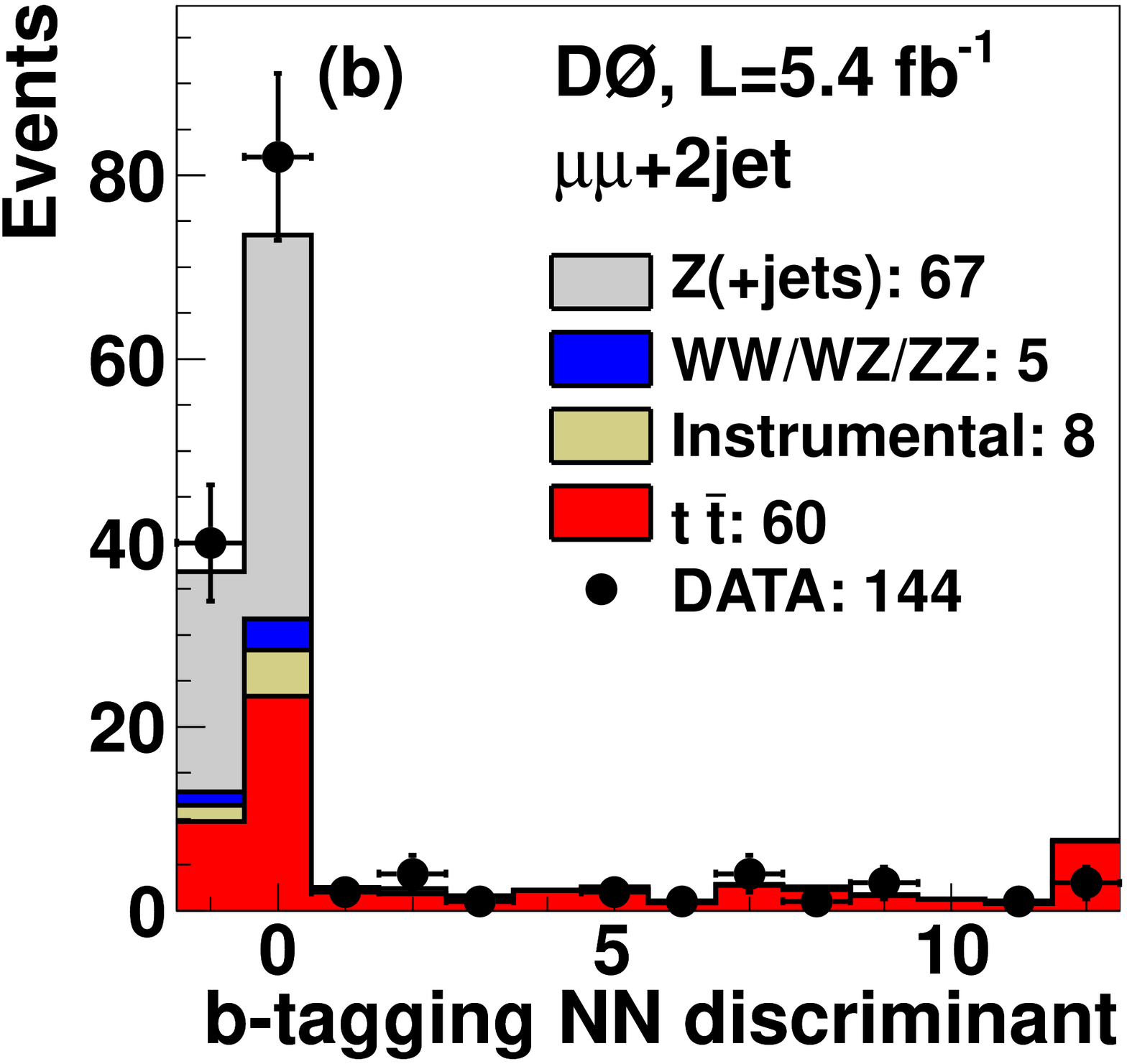}
\includegraphics[width=.4\textwidth]{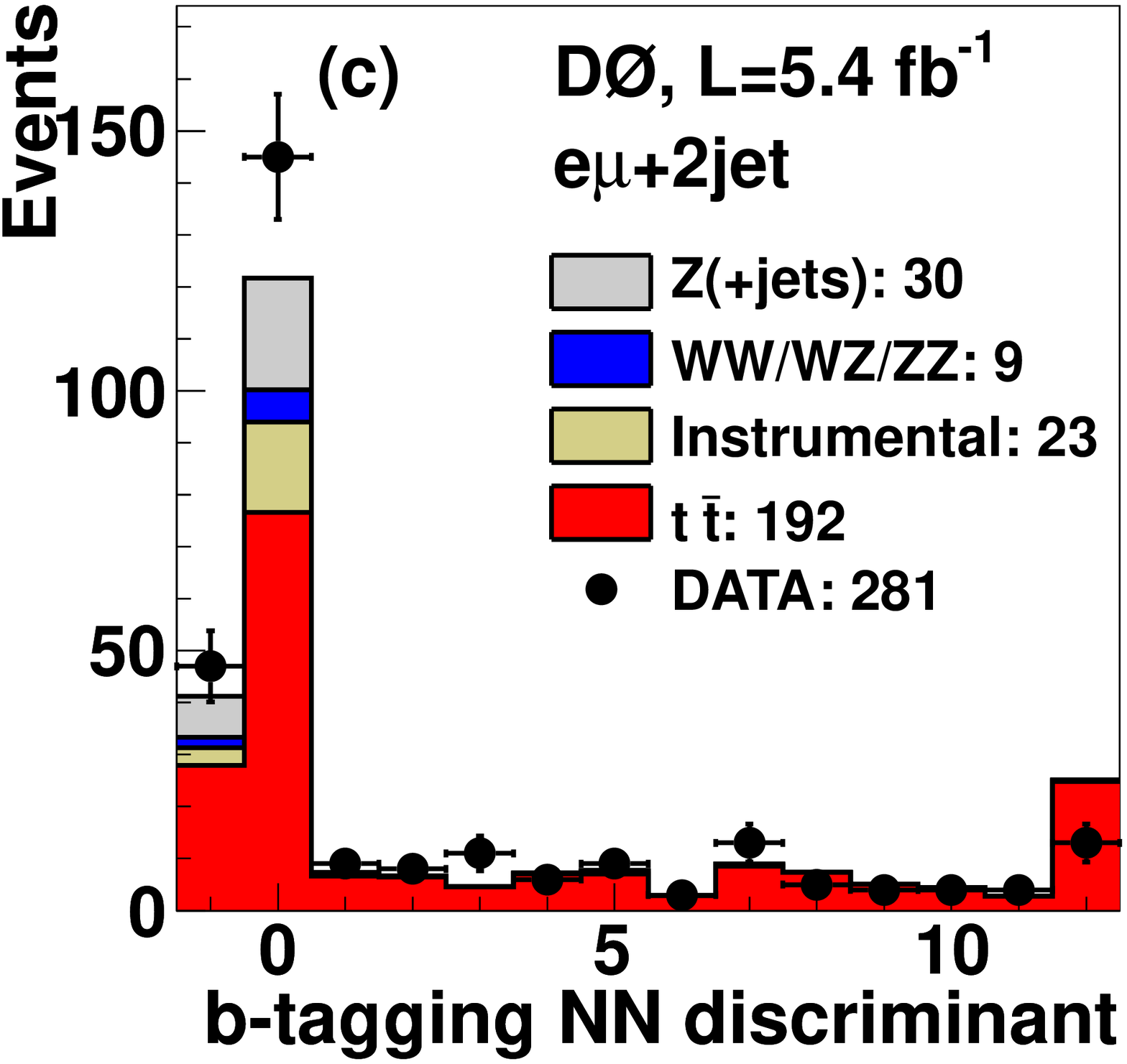}
\hfill
\includegraphics[width=.4\textwidth]{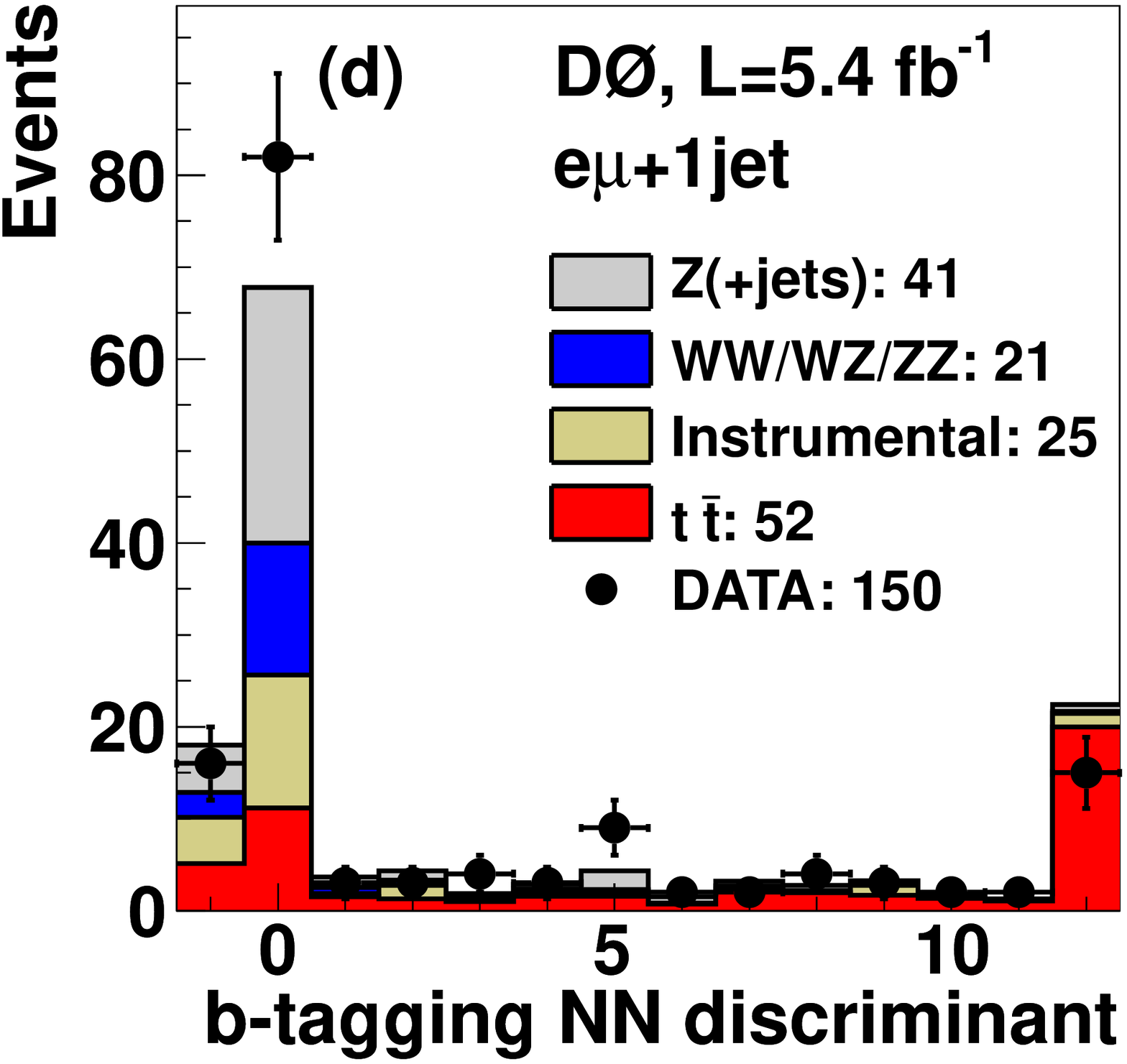}
\caption{
\label{fig:nn}
Expected and observed distributions for the smallest 
$b$-tagging NN discriminant output of the two leading jets
for the (a) $ee$ + 2 jet channel, (b) $\mu\mu$ + 2 jet channel, 
(c) $e\mu$ + 2 jet channel, and (d) $e\mu$ + 1 jet channel.
The \ttbar\ signal is normalized to the SM cross section (7.45 pb).
The $x$~axis represents the NN output non-uniformly mapped to  14 bins.
The bin with central value 0 represents the lowest probability for a jet to be produced by a $b$ quark.
The bin with value 12 represents the highest probability.
The bin with value $-1$ represents the jets which do not satisfy the requirements to enter the 
NN computation (non-taggable jets).}
\end{figure*}

The number of predicted background events as well as the expected number of signal 
events in the four channels are shown in Table~\ref{tab:yield}.
The $\ttbar$ events have two $b$ quark jets in the final state, but most of the 
background events have jets produced by light quarks or gluons.
In order to achieve  a better separation between signal and background
when measuring the cross section, we use the distribution of the smallest
of the two $b$-tagging NN discriminants of the two leading jets.
If a jet does not satisfy the requirements to enter the NN computation (non-taggable jet), a value of -1 is assigned to it.
These NN discriminant distributions for the different channels are shown in Fig.~\ref{fig:nn} 

We measure the \ttbar\ cross section \sigmattbar\ by simultaneously fitting the NN distributions in the four channels
and maximizing the likelihood function
\begin{equation}
\mathcal{L} = 
 \prod_{i}\prod_{j} P[n_{ij},\mu_{ij}(\sigmattbar)]\ ,
\label{eq:fitlhood}
\end{equation}
where $i$ runs over the channels and $j$ over the bins of the  NN distribution, and 
$P[n, \mu(\sigmattbar)]$ is the Poisson probability function to observe 
$n$ events when $\mu(\sigmattbar)$ events are expected.
%%%%%%%%%%%%%%%%%% BACKGROUND-SIGNAL %%%%%%%%%%%%%%%%%%
\begin{table*}
\caption{Numbers of expected and observed events
assuming the SM \ttbar\ cross section for a top quark mass of $m_t=172.5$~GeV (7.45 pb).
Expected numbers of events are shown with their systematic uncertainties.
The uncertainty on the ratio between observed and expected numbers of events 
takes into account the statistical uncertainty in the observed number of events ($N_{\rm obs}$) and the systematic
uncertainty in the expected number of events ($N_{\rm exp}$).
\label{tab:yield}
}
\begin{tabular}[t]{l|cccc|ccc}
\hline
Channel 
&$Z\to \ell\ell$ 
&Diboson 
& \parbox{2.0cm}{Instrumental background}
& $t\bar{t}\to \ell\bar{\ell} b\bar{b} \nu\bar{\nu}$  
& \parbox{1.8cm}{$N_{\rm exp}$}  
& \parbox{1.6cm}{$N_{\rm obs}$} 
& \parbox{1.2cm}{$\frac{\text{Observed}}{\text{Expected}}$} 
\\[6pt] \hline
$ee$+2jet & $12.6\pm 2.0$ & $3.0\pm 0.4$ &  -  & $45.6\pm 5.3$ & $61.1\pm 7.1$ & 74 & $1.21\pm0.20$\\ \hline
$\mu\mu$+2jet & $67.3\pm 9.7$ & $5.1\pm 0.7$ & $7.6\pm 1.2$ & $59.8\pm 6.6$ & $139.8\pm 15.7$ & 144 & $1.03\pm0.14$\\ \hline
$e\mu$+2jet & $30.3\pm 4.2$ & $8.6\pm 1.2$ & $22.7\pm 8.6$ & $191.5\pm 18.8$ & $253.1\pm 24.3$ & 281 & $1.11\pm0.13$\\ \hline
$e\mu$+1jet & $40.9\pm 4.8$ & $20.7\pm 2.4$ & $25.3\pm 10.5$ & $52.1\pm 9.4$ & $139.0\pm 16.5$ & 150 & $1.08\pm0.16$\\ \hline
\end{tabular}
\end{table*}

%%%%%%%%%%%%%%%%%% RESULTS %%%%%%%%%%%%%%%%%%
\section{Results and Uncertainties}

The main systematic uncertainties for the measurement of the \ttbar\ cross section are described in the following.
A $6.1 \%$ uncertainty~\cite{lumi} directly affects the cross section measurement because of the luminosity uncertainty 
but also the expected numbers of \Z\ and diboson background events. 
Uncertainties in lepton identification efficiencies are determined by evaluating
possible sources of bias in the data driven method used for the
efficiency measurements and the possible impact of data/MC
differences in \Z$\to\ell\ell$ events.
Uncertainties in the lepton energy resolution are determined by comparing the width of the 
$Z$ boson invariant mass distributions in data and MC. 

The uncertainty in the relative JES between data and MC for light quark jets
has been evaluated by shifting the jets in MC by their corresponding JES uncertainty.
The uncertainty on the difference between the light and $b$ quark JES
(1.8\%) is estimated by propagating the difference in the single pion response between data and MC to 
the MC JES for $b$ quark jets.
Jet energy resolution uncertainties are estimated 
by comparing the resolutions measured in \Z+jets events in data and in MC.
The uncertainty on the jet identification efficiency is estimated 
by comparing the efficiencies measured in dijet events for  data and MC.
The $b$ quark identification uncertainties include uncertainties in the probability
of tagging a $b$ quark jet, the probability of tagging a light quark jet or gluon,
and the probability for a jet not to be taggable~\cite{btagging}. 

To estimate the uncertainty in the trigger efficiency, we use events selected with the same criteria
as the \ttbar\ signal but without jet requirements. 
In all four channels this selection is dominated by \Z\ events.
We compute the ratio of the expected and observed number of events for two cases:
when both leptons are allowed to fire the trigger or 
when only one lepton is allowed to fire the trigger.
The difference in these ratios is used to estimate
the uncertainty on the trigger efficiency. 

Several uncertainties on the signal modeling are considered.
The effects of higher order corrections and the hadronization modeling 
are estimated as the difference in 
signal efficiencies using the default \alpgen + \pythia\ simulation and 
using events generated with  the \mcatnlo~\cite{Frixione:2006gn} + \herwig~\cite{Corcella:2000bw} simulation.
The uncertainty coming from color reconnection is evaluated by 
comparing the \ttbar\ efficiency using \pythia\ v6.4 tune Apro and \pythia\ v6.4 tune ACRpro~\cite{cr}.
The uncertainty on initial (ISR) and final (FSR) state radiation is evaluated
by varying the  ISR/FSR parameters in \pythia\  up to 20\% and
evaluating the change in the signal efficiency.
The uncertainty due to PDFs is estimated by reweighting the signal efficiency to the CTEQ6.1M PDFs~\cite{Nadolsky:2008zw} 
and looking at the efficiency variation for each eigenvector set that define the CTEQ6.1M uncertainty range.
The uncertainty due to the simulation of $b$ quark fragmentation 
is assigned to be the difference between tuning the parameters of the $b$ quark fragmentation function to LEP
or SLD data~\cite{bfrag}.

The uncertainty in the background normalization includes the theoretical uncertainties in the cross section and 
the uncertainty due to the correction for the $Z$ boson \pt modeling.
We also take into account an  uncertainty due to the limited statistics of the signal and background templates 
of the NN discriminant.
For the following systematic uncertainties, we take into account
effects that change the shape of the differential distribution of 
the $b$-tagging NN output discriminant:
jet energy scale, jet resolution, jet identification, and $b$ quark identification uncertainties.

Maximizing the likelihood function in Eq.~\ref{eq:fitlhood} and using the above systematic uncertainties, 
we measure the cross section assuming a top quark mass $m_t=172.5$~GeV and  find
\begin{equation} \label{eq:xsecstandard}
\sigmattbar = 8.05^{+0.50}_{-0.48} \ ({\rm stat}) ^{+1.05}_{-0.97} \ ({\rm syst})\ {\rm pb.}
\end{equation}

In order to reduce the influence of systematic uncertainties on the cross section measurement,
in the following we use nuisance parameters~\cite{nuisance}
to constrain the overall uncertainty using the data NN output distribution itself.
Using this technique, the likelihood~(Eq.~\ref{eq:fitlhood}) is modified,
\begin{equation} 
\mathcal{L} = 
\prod_{i}\prod_{j}P[n_{ij},\mu_{ij}(\sigmattbar, \nu_k)]
\ \prod_{k}\mathcal{G}(\nu_k; 0, SD),
\label{eq:dilepll}
\end{equation}
where
${\cal G}(\nu_k;0,{\rm SD})$ denotes the Gaussian probability density with mean at
zero and width corresponding
to one standard deviation (SD) of the considered systematic uncertainty.
Correlations of systematic uncertainties between channels and 
between the different samples are naturally taken into account by assigning the same nuisance parameter 
to the correlated systematic uncertainties. 
In Eq.~\ref{eq:dilepll}, the free parameters of the fit are $\nu_k$ and \sigmattbar.

As can be seen from~Eq.~\ref{eq:xsecstandard}, the systematic uncertainties are the limiting uncertainties in the precision 
of the $t\bar{t}$ cross section measurement. 
Varying  the systematic uncertainties and constraining them with data 
can therefore improve the measurement.  
Using nuisance parameters 
we find an overall improvement of the uncertainty of 20\%
and reach a relative 
precision of $11\%$ in the \ttbar\ cross section:
\[ 
\sigmattbar = \fullresultll\ {\rm pb}.
\]

The uncertainties on the $t\bar{t}$ cross section are summarized in Table~\ref{tab:syst}. 
For each category of systematic uncertainty listed in Table~\ref{tab:syst}, 
the corresponding nuisance parameters are set to their fitted value, and shifted by the uncertainty on the fit.
In the columns ``$+\sigma$'' and ``$-\sigma$,'' the positive and negative systematic uncertainties on the measured cross section for each category are listed. 

\begin{table*}[ht]
\newcommand\T{\rule{0pt}{2.6ex}}
\newcommand\B{\rule[-1.2ex]{0pt}{0pt}}
\begin{center} 
\begin{minipage}{7.0 in}
\caption{Breakdown of uncertainties on the $t\bar{t}$ cross sections 
in the $\ell\ell^\prime$ channel and for the combined $\ell\ell^\prime$ and $\ell j$ measurement 
using the nuisance parameter technique. 
The $\pm \sigma$ give the impact on the measured cross section when the nuisance parameters describing the considered category are 
shifted by $\pm 1~SD$ from their fitted mean.
See text for further details.
}\label{tab:syst}
\setlength{\tabcolsep}{5pt}
{
\renewcommand{\arraystretch}{1.1} 
\begin{tabular}{l c  c| c  c} \hline
&\multicolumn{2}{c|}{$\ell \ell^\prime$} & \multicolumn{2}{c}{$\ell \ell^\prime$+$\ell j$} \\ \hline
Source                                  & $+\sigma$ [pb]  & $-\sigma$ [pb] & $+\sigma$ [pb] & $-\sigma$ [pb] \\[1pt] \hline
Statistical 	 			& +0.50 & $-0.48$  & +0.20  & $-0.20$  \\
\hline
Muon identification 			& +0.11 & $-0.11$  & +0.07  & $-0.06$  \\
Electron identification and smearing 	& +0.24	& $-0.23$  & +0.13  & $-0.13$  \\
Signal modeling 			& +0.34	& $-0.33$  & +0.16  & $-0.06$  \\
Triggers 				& +0.19	& $-0.19$  & +0.05  & $-0.05$  \\
Jet energy scale 			& +0.13	& $-0.12$  & +0.04  & $-0.04$  \\
Jet reconstruction and identification 	& +0.21	& $-0.20$  & +0.12  & $-0.09$  \\
$b$-tagging 				& +0.06	& $-0.06$  & +0.16  & $-0.14$  \\
Background normalization 		& +0.29	& $-0.27$  & +0.11  & $-0.10$  \\
$W$+HF fraction				& -     & -        & +0.12  & $-0.04$  \\
Instrumental background 		& +0.18	& $-0.17$  & +0.05  & $-0.04$  \\
Luminosity 				& +0.57	& $-0.51$  & +0.48  & $-0.43$  \\
Other 					& +0.10	& $-0.10$  & +0.06  & $-0.06$  \\
Template statistics 			& +0.08	& $-0.08$  & +0.04  & $-0.04$  \\
\hline
 \end{tabular} 
 }
\end{minipage}
\end{center} 
\end{table*} 

We  combine this measurement  with the cross section measurement in the non-overlapping $\ell j$
channel~\cite{d0-top-ljets-5.3fb-1} using the same nuisance parameter approach
and taking correlations between common systematics uncertainties into account. 
%The $\ell j$ channel includes events where the leptonically decaying $W$ boson decays into an electron and a neutrino, a muon and a neutrino, or leptonically decaying taus and a neutrino. 
In the $\ell j$ channel, the events are separated into events with three or at least four jets, of which  zero, 
one, or at least two jets are $b$-tagged. 
In events that have three or four jets but no $b$-tagged jets or events with three jets and one $b$-tagged jet, we use a
topological discriminant to improve the separation of signal and background.
In~\cite{d0-top-ljets-5.3fb-1}, the separation into these channels and application of topological 
methods is referred to as the combined method. 
For this combination, we do not simultaneously fit the heavy flavor fraction for $W$+jet processes ($W$+HF) 
in the $\ell j$ channel as was done in~\cite{d0-top-ljets-5.3fb-1}, making it unnecessary to 
use $\ell j$ events with only two jets. With this change compared to~\cite{d0-top-ljets-5.3fb-1}, the
measured $\ell j$ \ttbar\ cross section is
\[
\sigmattbar =  \fullresultlj\ {\rm pb}.
\]

The combination of the measurements in the  dilepton and lepton + jet final states is done by maximizing the 
product of the likelihood function for the $\ell\ell^\prime$ and $\ell j$ channels, which yields
\[
\sigmattbar = \fullresult\ {\rm pb}
\]
for $ m_t = 172.5$~GeV.
This combination has a relative precision of $8\%$ and represents an improvement of about $12\%$
relative to the $\ell j$ cross section measurement alone. The uncertainties for this combined 
measurement are summarized in Table~\ref{tab:syst}. 

\begin{figure}
\centerline{\includegraphics[width=0.5\textwidth, height=0.5\textwidth]
{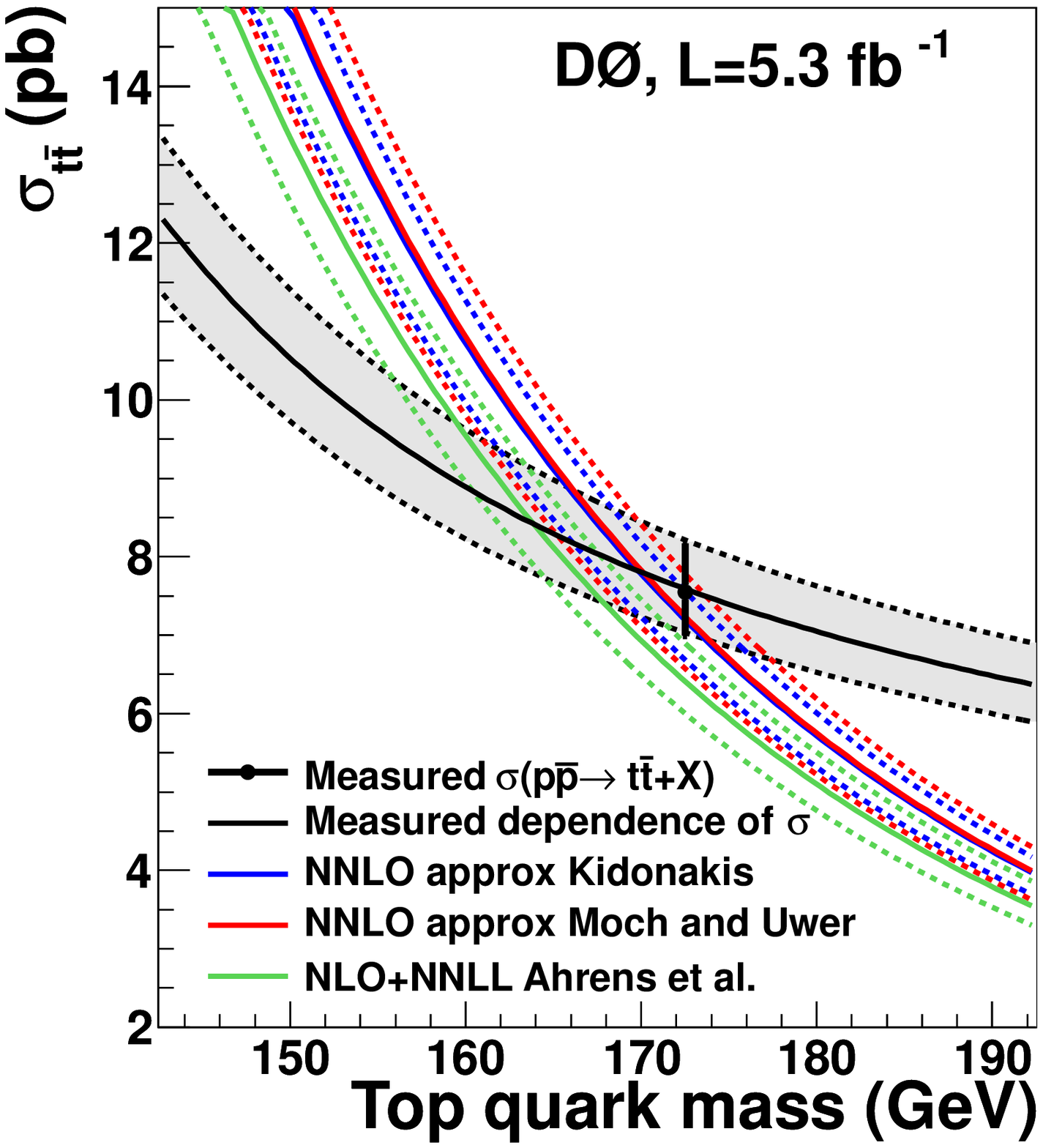}}
\caption{Dependence of the experimental and theoretical~\cite{SMtheory_A,SMtheory_M,SMtheory_K} 
\ttbar\ cross sections on the top quark mass. The colored dashed lines represent the uncertainties for all three theoretical calculations
from the choice of the PDF and the renormalization and factorization scales (added quadratically).
The data point shows the combined $\ell\ell^\prime$ and $\ell j$ cross section measurement for $m_t=172.5$~GeV, the black curve is
experimental \ttbar\ cross section as a function of $m_t$, and the gray band corresponds to the total experimental uncertainty.}
\label{fig:massdep}
\end{figure}

Due to acceptance effects, the \ttbar\ efficiency depends on the assumed \mt\ in the MC. 
We extract the \ttbar\ cross section using simulated \ttbar\ events 
with different values of $m_t$.
The resulting cross sections can be fitted with the following functional form: 
\begin{eqnarray}
\lefteqn{\sigmattbar(m_t)  =   \frac{1}{m^4_t} [ a + b (m_t-170\ {\rm GeV})} \nonumber \\
& & \mbox{}+ c (m_t-170\ {\rm GeV})^2 +  d (m_t-170\ {\rm GeV})^3  ]~,
\end{eqnarray}
with $a = 6.5178 \times 10^{9}$ GeV$^4$, $b = 7.884 \times 10^{7}$ GeV$^3$, 
$c = 9.3069 \times 10^{5}$ GeV$^2$, and $d = -2.42 \times 10^{3}$ GeV 
and where $\sigma_{t\bar{t}}$ and $m_t$ are in pb and GeV, respectively.
The relative uncertainty on the cross section for different mass points is the same
as the one obtained for $m_t=172.5$~GeV.
Figure~\ref{fig:massdep} shows 
this parameterization for the measurement as a function of top quark mass 
together with approximate NNLO computations~\cite{SMtheory_A,SMtheory_M,SMtheory_K}.

\section{Conclusion}
In this Letter we presented an updated measurement of the \ttbar\ production 
cross section in the dilepton final state using \lumi\ of integrated luminosity.
This cross section measurement yields \sigmattbar=\fullresultll~pb and has a relative precision of $^{+12\%}_{-11\%}$.
It is currently the most precise  measurement of the \ttbar\ cross section in the dilepton channel.
Combining this measurement with our result in the lepton + jets 
channel~\cite{d0-top-ljets-5.3fb-1} yields \fullresult~pb which corresponds to a relative precision of 8\%. 
This measurement is in good agreement with the SM prediction.

We thank the staffs at Fermilab and collaborating institutions,
and acknowledge support from the
DOE and NSF (USA);
CEA and CNRS/IN2P3 (France);
FASI, Rosatom and RFBR (Russia);
CNPq, FAPERJ, FAPESP and FUNDUNESP (Brazil);
DAE and DST (India);
Colciencias (Colombia);
CONACyT (Mexico);
KRF and KOSEF (Korea);
CONICET and UBACyT (Argentina);
FOM (The Netherlands);
STFC and the Royal Society (United Kingdom);
MSMT and GACR (Czech Republic);
CRC Program and NSERC (Canada);
BMBF and DFG (Germany);
SFI (Ireland);
The Swedish Research Council (Sweden);
and
CAS and CNSF (China).
%

%%%%%%%%%%%%%%%%%%%%%%%%%%%%%%%%%%%%%%%%%%%%%%%%%%%%%%%%%%


\begin{thebibliography}{99}

\bibitem{SMtheory_A}
  V.~Ahrens, A.~Ferroglia, M.~Neubert, B.~D.~Pecjak, and L.~L.~Yang,
  %``Renormalization-Group Improved Predictions for Top-Quark Pair Production at
  %Hadron Colliders,''
  \Journal{J.~High~Energy~Phys.}{09}{097}{2010};
  %[arXiv:1003.5827 [hep-ph]].
  %%CITATION = JHEPA,1009,097;%%
  %\cite{Ahrens:2010mj}
  V.~Ahrens, A.~Ferroglia, M.~Neubert, B.~D.~Pecjak, and L.~L.~Yang,
  %``Top-Quark Pair Production Beyond Next-to-Leading Order,''
  \Journal{Nucl.\ Phys.\ Proc.\ Suppl.}{205--206}{48}{2010}.
  %[arXiv:1006.4682 [hep-ph]].
  %%CITATION = NUPHZ,205-206,48;%%


\bibitem{SMtheory_M}
  S.~Moch and P.~Uwer, 
  \Journal{Phys.~Rev.~D}{78}{034003}{2008}; 
  U.~Langenfeld, S.~Moch, and P.~Uwer,
  %``Measuring the running top-quark mass,''
  \Journal{Phys.~Rev.~D}{80}{054009}{2009}.
  %[arXiv:0906.5273 [hep-ph]].
  %%CITATION = PHRVA,D80,054009;%%


\bibitem{SMtheory_K} 
  N.\ Kidonakis and R.\ Vogt, \Journal{Phys.~Rev.~D}{68}{114014}{2003};
  N.~Kidonakis,
  %``Next-to-next-to-leading soft-gluon corrections for the top quark cross
  %section and transverse momentum distribution,''
  \Journal{Phys.~Rev.~D}{82}{114030}{2010}.
  %[arXiv:1009.4935 [hep-ph]].
  %%CITATION = PHRVA,D82,114030;%%


\bibitem{d0-topmass-fromxsection}
V.~M.~Abazov {\it et al.} (\dzero\ Collaboration), \Journal{\PLB}{679}{177}{2009}.

\bibitem{d0-top-charged-higgs}
V.~M.~Abazov {\it et al.} (\dzero\ Collaboration), \Journal{Phys.~Rev.~D}{80}{071102}{2009}.

\bibitem{d0-top-ljets-5.3fb-1}
V.~M.~Abazov {\it et al.} (\dzero\ Collaboration), \Journal{Phys.~Rev.~D}{84}{012008}{2011}.

\bibitem{d0-top-ljets}
%V.~M.~Abazov {\it et al.} (\dzero\ Collaboration), \Journal{\PLB}{626}{35}{2005};
%V.~M.~Abazov {\it et al.} (\dzero\ Collaboration), \Journal{\PLB}{626}{45}{2005};
%V.~M.~Abazov {\it et al.} (\dzero\ Collaboration), \Journal{Phys.~Rev.~D}{74}{112004}{2006};
%V.~M.~Abazov {\it et al.} (\dzero\ Collaboration), \Journal{Phys.~Rev.~D}{76}{092007}{2007};
V.~M.~Abazov {\it et al.} (\dzero\ Collaboration), \Journal{\PRL}{100}{192004}{2008}.

\bibitem{d0-top_alljets-1fb-1}
V.~M.~Abazov {\it et al.} (\dzero\ Collaboration), \Journal{Phys.~Rev.~D}{76}{072007}{2007};
V.~M.~Abazov {\it et al.} (\dzero\ Collaboration), \Journal{Phys.~Rev.~D}{82}{032002}{2010}.

\bibitem{d0-top-dilepton-1fb}
V.~M.~Abazov {\it et al.} (\dzero\ Collaboration), \Journal{\PLB}{626}{55}{2005}.

\bibitem{CDF-dilepton}
%D.~Acosta {\it et al.} (CDF Collaboration), \Journal{\PRL}{93}{142001}{2004};
%A.~Abulencia {\it et al.} (CDF Collaboration), \Journal{Phys.~Rev.~D}{78}{012003}{2008};
%T.~Aaltonen {\it et al.} (CDF Collaboration), \Journal{Phys.~Rev.~D}{79}{112007}{2009};
T.~Aaltonen {\it et al.} (CDF Collaboration), \Journal{Phys.~Rev.~D}{82}{052002}{2010}.

\bibitem{atlas-top}
G.~Aad {\it et al.} (ATLAS Collaboration), 
\Journal{Eur.~Phys.~J.~C}{71}{1577}{2011}. 

\bibitem{cms-top}
V.~Khachatryan {\it et al.} (CMS Collaboration), \Journal{\PLB}{695}{424}{2011}.

\bibitem{d0-detector}
V.~M.~Abazov {\it et al.} (\dzero\ Collaboration), \Journal{\NIMA}{565}{463}{2006}.

\bibitem{lumi} T.~Andeen  {\it et al.}, FERMILAB-TM-2365 (2007).

\bibitem{geant} 
  R.\ Brun, F.\ Carminati, CERN Program Library Long Writeup W5013, 
  1993 (unpublished).  

\bibitem{RunIIcone}
G.~C.~Blazey {\it et al.}, in
     {\sl Proceedings of the Workshop:
     ``QCD and Weak Boson Physics in Run II,''}
     edited by U.~Baur, R.~K.~Ellis, and D.~Zeppenfeld
     (Fermilab, Batavia, IL, 2000) p.~47,
     arXiv:hep-ex/0005012;
     see Sec. 3.5 for details.
   
%\bibitem{d0-l0}
%R.~Angstadt {\it et al.} (\dzero Collaboration), \Journal{\NIMA}{622}{298}{2010}

\bibitem{btagging} 
  V.~M.~Abazov {\it et al.}  (\dzero\ Collaboration) \Journal{\NIMA}{620}{490}{2010}.

\bibitem{d0-top-dilepton-400pb}
V.~M.~Abazov {\it et al.} (\dzero\ Collaboration), \Journal{Phys.~Rev.~D}{76}{052006}{2007}.
   
\bibitem{Alpgen} 
  % \cite{Mangano:2002ea} 
  M.~L.~Mangano, M.~Moretti, F.~Piccinini, R.~Pittau, and A.~D.~Polosa, 
  % ``ALPGEN, a generator for hard multiparton processes in hadronic 
  % collisions,'' 
  \Journal{J.~High~Energy~Phys.}{07}{001}{2003}. 
  % [arXiv:hep-ph/0206293]. 
  %% CITATION = JHEPA,0307,001;%% 
   
\bibitem{Pythia} 
  % \cite{Sjostrand:2006za} 
  T.~Sj\"ostrand, S.~Mrenna, and P.~Z.~Skands, 
  % ``PYTHIA 6.4 Physics and Manual,'' 
   \Journal{J.~High~Energy~Phys.}{05}{026}{2006}.
  % [arXiv:hep-ph/0603175]. 
  %% CITATION = JHEPA,0605,026;%% 

%\bibitem{PDF}
%J. Pumplin {\it et al.} (CTEQ Collaboration),
%\Journal{J.~High~Energy~Phys.}{07}{012}{2002}.

\bibitem{Nadolsky:2008zw}
  P.~M.~Nadolsky {\it et al.},
  %``Implications of CTEQ global analysis for collider observables,''
  \Journal{Phys.~Rev.~D}{78}{013004}{2008}.
%  [arXiv:0802.0007 [hep-ph]].

\bibitem{fewz}
 R.~Gavin, Y.~Li, F.~Petriello, and S.~Quackenbush,
  %``FEWZ 2.0: A code for hadronic Z production at next-to-next-to-leading
  %order,''
  arXiv:1011.3540 [hep-ph].

\bibitem{Ellis:2006ar}
  R.~K.~Ellis,
  %``An update on the next-to-leading order Monte Carlo MCFM,''
  \Journal{Nucl.~Phys.~Proc.~Suppl.}{160}{170}{2006}.

\bibitem{zptrw}
  V.~M.~Abazov {\it et al.} (\dzero\ Collaboration),
  %``Measurement of the normalized $Z/\gamma^* -> \mu^+\mu^-$ transverse momentum distribution in $p\bar{p}$ collisions at $\sqrt{s}=1.96$ TeV,''
  \Journal{\PLB}{693}{522}{2010}.

\bibitem{Frixione:2006gn}
  S.~Frixione and B.~R.~Webber, \Journal{J.~High~Energy~Phys.}{0206}{029}{2002},
  S. Frixione, P. Nason, B.R. Webber, \Journal{J.~High~Energy~Phys.}{0308}{007}{2003}.

\bibitem{Corcella:2000bw}
  G.~Corcella, I.~G.~Knowles, G.~Marchesini, S.~Moretti, K.~Odagiri, P.~Richardson, M.~H.~Seymour, B.~R.~Webber,
  %``HERWIG 6: An Event generator for hadron emission reactions with interfering gluons (including supersymmetric processes),''
  \Journal{J.~High~Energy~Phys.}{01}{010}{2001}.
%   arXiv:hep-ph/0011363.

\bibitem{cr}
  A.~Buckley, H.~Hoeth, H.~Lacker, H.~Schulz and J.~E.~von Seggern,
  %``Systematic event generator tuning for the LHC,''
  \Journal{Eur.~Phys.~J.~C}{65}{331}{2010};
%  [arXiv:0907.2973 [hep-ph]];
  %%CITATION = EPHJA,C65,331;%%

  P.~Z.~Skands and D.~Wicke,
  %``Non-perturbative QCD effects and the top mass at the Tevatron,''
  \Journal{Eur.~Phys.~J.~C}{52}{133}{2007}.
%  [arXiv:hep-ph/0703081].
  %%CITATION = EPHJA,C52,133;%%



\bibitem{bfrag}
Y. Peters, K. Hamacher and D. Wicke, FERMILAB-TM-2425-E.

\bibitem{nuisance} 
  P.~Sinervo,
  %``Definition and treatment of systematic uncertainties in high energy
  %physics and astrophysics,''
 In {\it Proceedings of PHYSTAT 2003: ``Statistical Problems in Particle Physics, Astrophysics, and Cosmology''},
 edited by L. Lyons, R. Mount, and R. Reitmeyer (SLAC, Menlo Park, CA, 2003), p. 122, SLAC-R-703.

\end{thebibliography}
\end{document}